# Hokusai — Sketching Streams in Real Time

**Sergiy Matusevych**
Yahoo! Research
sergiy.matusevych@gmail.com

**Alexander J. Smola**
Google Inc.
alex@smola.org

**Amr Ahmed**
Yahoo! Research
amahmed@cs.cmu.edu

## Abstract

We describe 北斎 Hokusai, a real time system which is able to capture frequency information for streams of arbitrary sequences of symbols. The algorithm uses the Count-Min sketch as its basis and exploits the fact that sketching is linear. It provides real time statistics of arbitrary events, e.g. streams of queries as a function of time. We use a factorizing approximation to provide point estimates at arbitrary (time, item) combinations. Queries can be answered in constant time.

## 1 Introduction

Obtaining frequency information of data streams is an important problem in the analysis of sequence data. Much work exists describing how to obtain highly efficient frequency counts of sequences at any given time. For instance, the Count-Min [6] sketch is capable of providing $\epsilon$ accuracy with failure probability $\delta$ in $O(\log \delta/\epsilon)$ space. Even better guarantees are possible for the space-saver sketch [12], albeit at the expense of having to keep a dictionary of tokens and a considerably more computationally expensive data structure (we need an ordered list rather than just a flat array in memory). The key point is that sketching algorithms can be queried at any time to return the total number of symbols of any given type seen so far.

While highly desirable in its own right it does not solve the following: we would like to know how many symbols were observed at some time in the past. For instance, we may want to know how many queries of "Britney Spears" were carried out, say at noontime last year on Easter Sunday. Clearly we could address this problem by brute force, that is by pre-processing the logfiles. Query languages such as Pig [14] are well suited to generating such summary statistics, albeit unable to retrieve it in *real time*.

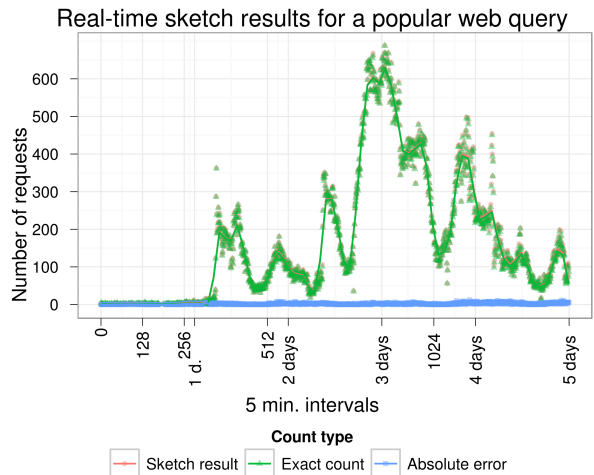

Figure 1: Real-time results for the item aggregation sketch and web query (*"gigi goyette"*) that was popular at the time of our study. Points represent actual results; lines are LOESS smoothed with span 0.1. The plot shows the exact time query started to gain popularity and peaked, as well as daily fluctuations of search traffic. Estimate and real traffic are virtually indistinguishable and the error is minimal.

It is the latter that we focus on in this paper. We present an algorithm capable of retrieving *approximate* count statistics in real time for any given point or interval in time without the need to peruse the original data. We will see that the work required for retrieving such statistics is $O(\log t)$ for lookup and $O(t)$ for decoding (here $t$ is the length of the requested sequence) and moreover that the answers are exact under the assumption of statistical independence. Experiments show that even if this assumption is violated, as it typically is in practice, the statistics still match closely the exact data. We use a number of blocks:

- The Count-Min sketch serves as the basic data aggregator. This is desirable since it has the property of being linear in the data stream. That is, if we sketch time periods $T$ and $T'$, then the sketch of $T \cup T'$ is given by the sum over the two individual sketches. This makes it easy to aggregate time intervals which are powers of 2.
- We exploit the fact that the Count-Min sketch is linear in the resolution of the hash function. That is, a sketch using a lower resolution hash function is obtained by aggregating adjacent bins.
- We use the fact that for independent random variables the joint distribution can be characterized exactly by retaining only the marginals.
- We use a Markovian approximation for sequences.

Our implementation stores Count-Min sketches of both increasing time intervals in powers of 2, e.g. 1, 2, 4, 8, ... minutes length. Moreover we also store sketches of decreasing bit resolution. Finally, in order to accelerate computation, we store sketches which have both decreasing time and bit resolution. This provides a suitable normalization.

**Outline.** We begin by giving a brief overview over the Count-Min sketch [6] and how it can be parallelized and made fault tolerant. In Section 3 we describe the data structures used to aggregate the data in a linear fashion and show how reduction in the bit resolution of the hash function allows us to compute both time and keyspace marginals. Using approximate independence assumptions we can recover higher (time, item) resolution of the sketch over time. In Section 4 we apply sketches to $O(1)$ probability estimates in the context of probabilistic graphical models. Section 5 contains experimental results.

## 2  Count-Min Sketch

Algorithms for sketches of data streams aim at obtaining count statistics for observed items as they keep on arriving in an online fashion. That is, such algorithms typically allow one to assess how many items of a given kind are contained in the data stream. One of the most exciting algorithms for this purpose is the Count-Min sketch of [6], a generalization of the Bloom filter [3] and the count sketch [5], which generates linear data summaries *without* requiring storage of the keys. That is, the maximum amount of memory is allocated for storing count statistics rather than an auxiliary index.

Denote by $\mathcal{X}$ the domain of symbols in the data stream. The Count-Min sketch allocates a matrix of counters $M \in \mathbb{R}^{d \times n}$ initially all set to zero. Moreover, we require $d$ pairwise independent hash functions $h_1, \ldots, h_d : \mathcal{X} \to \{0, \ldots n-1\}$. For each item insertion $d$ counters, as determined by the hash functions and the key, are incremented. Retrieval works by taking the minimum of the associated counts (this mitigates errors due to collisions). Algorithm 1 comes with surprisingly strong guarantees. In particular [6] proves:

**Algorithm 1** Count-Min Sketch
  **insert**$(x)$:
  **for** $i = 1$ **to** $d$ **do**
    $M[i, h_i(x)] \leftarrow M[i, h_i(x)] + 1$
  **end for**

  **query**$(x)$:
  $c = \min \{M[i, h_i(x)] \text{ for all } 1 \leq i \leq d\}$
  **return** $c$

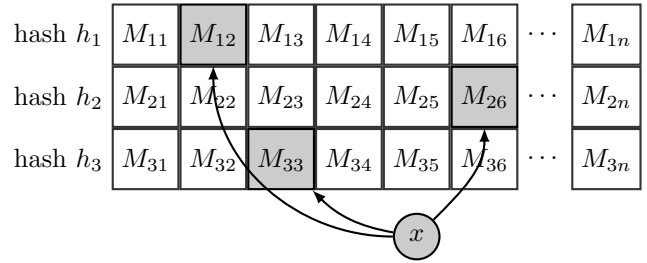

Figure 2: Item $x$ is inserted into each row of the Count-Min sketch (positions 2, 6, and 3 respectively).

**Theorem 1** *Assume that $n = \lceil \frac{e}{\epsilon} \rceil$ and $d = \lceil \log \frac{1}{\delta} \rceil$. Then with probability at least $1 - \delta$ the Count-Min sketch returns at any time an estimate $c_x$ of the true count $n_x$ for any $x$ which satisfies*

$$n_x \leq c_x \leq n_x + \epsilon \sum_{x'} n_{x'}. \qquad (1)$$

That is, the guarantee (1) is simply an additive bound on the deviation of counts. Tighter estimates can be obtained from $M$ by using an iterative procedure which goes beyond Algorithm 1. In particular, the Counter Braid iteration [11] provides guarantees of exact reconstruction with high probability provided that the number of stored symbols is sufficiently small and sufficiently well distributed. Note that reconstruction using [11] requires linear time, hence it is not applicable to our scenario of high throughput sketches. Furthermore, whenever the item distribution follows a power law, a tighter bound is available [7]. The following two corollaries are immediate consequences of the fact that the Count-Min sketch is *linear* in the amount of data and in the amount of memory available.

**Corollary 2** *Denote by $M_T$ and $M_{T'}$ the Count-Min sketches for the data stream observed at time intervals $T$ and $T'$. Then whenever $T \cap T' = \emptyset$ we have that $M_{T \cup T'} = M_T + M_{T'}$.*

This is analogous to Bloom filter hashes of the union of two sets — one simply takes the OR of both hashes. In our case the Boolean semiring is replaced by the semiring of integers (or real numbers) [4].

**Corollary 3** *Assume that $n = 2^b$ for $b \in \mathbb{N}$ and let $M_b$ be the summary obtained by the Count-Min sketch. In this case the summary $M_{b-1}$ obtained by a sketch $M_{b-1}[i,j] = M_b[i,j] + M_b[i, j+2^{b-1}]$ using hash functions $h_i^{b-1}(x) := h_i^b(x) \mod 2^{b-1}$.*

This is simply 'folding over' the first and second half of the Count-Min sketch on itself. These two corollaries mean that it is possible to increase the time intervals simply by aggregating sketches over shorter sub-intervals. Likewise, we may reduce the accuracy *after the fact* simply by dropping the most significant bits of $h$. This will become valuable when compressing the Count-Min sketch by reducing either temporal resolution or accuracy. Obviously as a consequence of this compression we double the error incurred by the sketch. The following chapters present algorithms to counteract this effect.

## 3 Aggregation

We want to obtain information regarding item frequency over an extended period of time. Assume that we are interested in this data at a resolution of 1 minute (the time scale is arbitrary — we just use 1 minute for the sake of concreteness). Even in the case where $M$ might be relatively small (in the order of 1 MB) we would suffer from memory overflow if we wanted to store 2 years (approximately $2^{20}$ minutes and hence 1 TB of data) in RAM to allow for fast retrieval.[1] Consequently we need to compress data.

### 3.1 Time Aggregation

One strategy is to perform binary aggregation on the time axis. That is, rather than retaining a 1 minute resolution for 2 years, we only retain a $2^m$ minute resolution for the last $2^m$ minutes. The rationale is that old observations have an exponentially decreasing value and consequently it is sufficient to store them at a concomitantly decreased precision. This is achieved by Algorithm 2. The basic idea is to keep aggregates $M^i$ for time intervals of length $\{1, 1, 2, 4, 8, \ldots, 2^m\}$ available. Note that the sequence contains its own cumulative sum, since $1 + \sum_{i=1}^{n-1} 2^i = 2^n$. To update $M^i$ whenever it covers the time span $[2^{i-1}, 2^i - 1]$ it suffices to sum over the terms covering the smaller segments.

**Theorem 4** *At $t$, the sketch $M^j$ contains statistics for the period $[t - \delta, t - \delta - 2^j]$ where $\delta = t \mod 2^j$.*

---

[1] We could store this data on disk. Many tools exist for offline parallel analytics which complement our approach. However, we are interested in online instant access.

---

**Algorithm 2** Time Aggregation
  **for all** $m$ **do do**
    Initialize Count-Min sketch $M^m = 0$
  **end for**
  Initialize $t = 0$ and $\bar{M} = 0$
  **while** data arrives **do**
    Aggregate data into sketch $\bar{M}$ for unit interval
    $t \leftarrow t + 1$ (increment counter)
    **for** $j = 0$ **to** $\operatorname{argmax} \{l \text{ where } i \mod 2^l = 0\}$ **do**
      $T \leftarrow \bar{M}$ (back up temporary storage)
      $\bar{M} \leftarrow \bar{M} + M^j$ (increment cumulative sum)
      $M^j \leftarrow T$ (new value for $M^j$)
    **end for**
    $\bar{M} \leftarrow 0$ (reset aggregator)
  **end while**

---

**Proof** The proof proceeds by induction. At time $t = 0$ this condition clearly holds since all counters are empty. Now assume that it holds for time $t > 0$. Denote by $j^* := \operatorname{argmax} \{l \text{ where } i \mod 2^l = 0\}$ the largest exponent of 2 for which $t$ is divisible by $2^{j^*}$. This is the last index which will become aggregated.

- For all $M^j$ with $j > j^*$ the condition holds if we increment $t \leftarrow t+1$ since the intervals are shifted by one time step without crossing any power of 2.
- For all $M^j$ with $j \leq j^*$ note that in the `for` loop we update all such counters by cumulative sums covering a consecutive contiguous sequence of shorter intervals via the cumulative sum counter $S$. In other words, all these intervals start from $t - \delta = 0$ again. ∎

**Lemma 5** *The amortized time required to update the statistics is $O(1)$ per time period regardless of $T$.*

**Proof** This follows from the fact in $2^{-i}$ of all time periods we need to do $i$ work. Hence the total amortized workload is $\sum_{i=1}^{\infty} i \cdot 2^{-i} = 2$. ∎

One of the nice side-effects of this procedure is that an "expensive" aggregation step which can involve up to $\log t$ many additions is always followed by a cheap update requiring only a single update (see Figure 3). This means that the update operations may be carried out as background threads in parallel to new updates, as long as they complete their first step within the update time period (this is trivially the case).

### 3.2 Item Aggregation

An alternative means of aggregating count statistics over time is to retain the full time resolution while sacrificing accuracy. That is, we can shrink the number of bins $n$ over time but we retain the unit time intervals. More specifically, we can halve the resolution each time the Count-Min sketch ages $2^n$ time steps. One of the advantages of this strategy is that the work required for aggregation is directly proportional to the remaining size of the bins (see Figure 3). We invoke Corollary 3 to halve the amount of space required successively. This yields Algorithm 3.

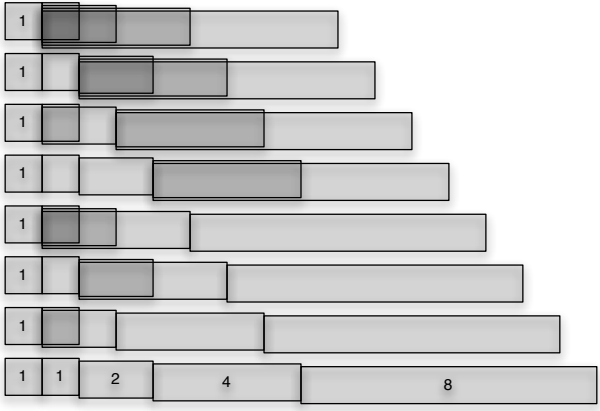

Figure 3: Time aggregation. We always insert into the leftmost aggregation interval. As time proceeds (top to bottom), new intervals are inserted from the left. Whenever an interval of length $2^n$ is $2^n$ steps old, it aggregates all data in $[0, 2^n]$ as its new value. The more shaded bins are contained in more than one interval.

**Algorithm 3** Item Aggregation
$t \leftarrow 0$ (initialize counter)
**while** data arrives **do**
  Receive aggregated data sketch $\bar{M}$ for unit interval
  $t \leftarrow t + 1$ (increment counter)
  $A^t \leftarrow \bar{M}$
  **for** $k = 1$ **to** $\lfloor \log_2 t \rfloor$ **do**
    **for all** $i, j$ **do** (reduce item resolution)
    $A^{t-2^k}[i, j] \leftarrow A^{t-2^k}[i, j] + A^{t-2^k}[i, j + 2^{m-k}]$
    Shrink $A^{t-2^k}$ to length $2^{m-k}$
  **end for**
**end while**

Note that as before this algorithm requires only $O(1)$ computational cost. Even better than before, this cost is now $O(1)$ for all steps rather than being an amortized (i.e. average) cost: at each time step there is one intervals of size $2^i$ with $2^i \leq n$ for each $i$ that needs halving. The cost of doing so is bounded by $\sum_{i=0}^{l} 2^i = 2^{l+1} - 1 < 2n$.

Moreover, the amount of storage required per time interval $[t - 2^i, t - 2^{i-1}]$ is constant since for every doubling of the interval size we halve the amount of storage required per sketch. This means that we can retain full temporal resolution, albeit at the expense of increasingly lose bounds on the item counts. In the extreme case where $2^m = n$ we end up with a sketch which tells us simply how many tokens we received at a given point in time but no information whatsoever regarding the type of the tokens.

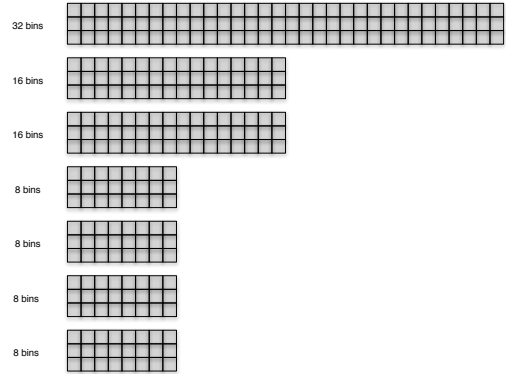

Figure 4: Item aggregation. Every $2^n$ with $n \in \mathbb{N}$ we halve the amount of bits used in the sketch. This allows us to sketch $T$ time steps in $O(\log T)$ space. Obviously having fewer storage bins means that we have less space available to store fine-grained information regarding the observed keys. However, for heavy hitters this is sufficient.

Note that by default the absolute error also doubles at every aggregation step (the scaling is slightly different for power law distributions). This means that after several iterations the accuracy of the sketch becomes very poor for low-frequency items while still providing good quality for the heavy hitters. We will exploit this as follows — for low frequency items we will use an interpolation estimate whereas for high frequency items we will use the aggregate sketch directly.

### 3.3 Resolution Extrapolation

We can take advantage of the following observation to improve our estimates: at some point in time we would like to obtain to-the-minute resolution of observations for arbitrary objects. In the extreme case we may have to-the-minute counts for *all* events and simultaneously highly specific counts for a *long* period of time. The basic idea in extrapolating counts is to use the fact that a joint distribution $p(x, t)$ over time and events can be recovered using its marginals $p(x)$ and $p(t)$ whenever $x$ and $t$ are independent of each other. In terms of counts this means that we estimate

$$\hat{n}_{xt} = \frac{n_x \cdot n_t}{n} \text{ where } n = \sum_t n_t = \sum_x n_x. \quad (2)$$

The quantities $n_x$ and $n_t$ are available (as upper bounds) via the Count-Min sketches $M^i$ and $A^j$ respectively. Finally, $n$ could be obtained at runtime by carrying out the summation over $t$ or $x$ respectively. However, this is too costly as it requires summing over $O(T)$ bins at time $T$. Instead, we compute a third set of aggregate statistics simultaneously to Algorithms 2 and 3 which performs both time and item aggregation.

---
**Algorithm 4** Item and Time Aggregation

$t \leftarrow 0$ (initialize counter)
**while** data arrives **do**
    Wait until item and time aggregation complete
    $t \leftarrow t+1$ (increment counter)
    $S = M^2$ (we only start combining at time 2)
    **for** $j = 1$ **to** $\mathrm{argmax}\left\{l \text{ where } i \bmod 2^l = 0\right\}$ **do**
        $S[i,j] \leftarrow S[i,j] + S[i, j+2^{m-j}]$
        Shrink $S$ to length $2^{m-j}$
        $T \leftarrow S$ (back up temporary storage)
        $S \leftarrow S + B^j$ (increment cumulative sum)
        $B^j \leftarrow T$ (new value for $B^j$)
    **end for**
**end while**

---

While individual estimates on $n_t$ and $n_x$ would allow for upper bounds via Theorem 1, and likewise a lower bound on $n$, we only obtain an approximation (we take a ratio between two upper bounds): we use (2) for each hash function separately and perform the min operation subsequently. At time $T$ the query for $n(x,t)$ requires:

$$\hat{n}(x,t) := \min_i \frac{M^{j^*}[i, h_i(x)] A^t[i, h_i^{m-j^*}(x)]}{B^{j^*}[i, h_i^{m-j^*}(x)]} \quad (3)$$

where $j^* := \lfloor \log_2(T-t) \rfloor$

We compute $j^*$ as the indicator for the most recent interval containing information pertaining $t$ after we have seen $T$ instants of data. The use of $h^{m-j^*}$ is needed to restrict the key range to the available array. As a consequence we obtain *estimates* of counts at full time and token resolution, albeit at decreasing accuracy as we look further into the past.

Since (3) is only an approximation of the true counts we use it only whenever the Count-Min estimate with reduced bit resolution is not accurate enough. Given that the accuracy decreases with $e2^{-b}$ where $2^b$ is the number of bins used in the sketch, we have the following variant of the sketching algorithm for queries: use the simplified item aggregate whenever the error is small enough. Otherwise switch to interpolation. Since we know that the absolute error of the Count-Min sketch is well controlled, we know that this strategy is self consistent for frequent items. See Algorithm 5 for details.

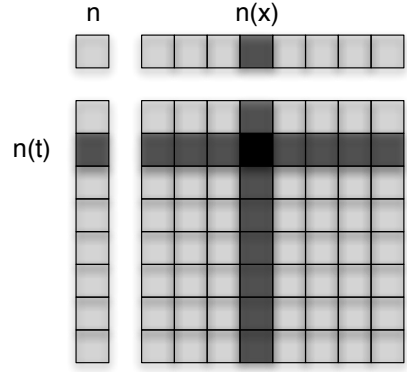

Figure 5: Resolution interpolation. We approximate $n(x,t)$ by $\frac{n(t)n(x)}{n}$ assuming independence, i.e. by approximating $p(a,b) \approx p(a) \cdot p(b)$. This is analogous to density estimation via copulas [13].

---
**Algorithm 5** Improved Interpolating Sketch

**query**$(x,t)$:
Obtain $\tilde{n}(x,t) := \min_i A^t[i, h_i^{m-j^*(x)}]$
**if** $\tilde{n}(x,t) > \frac{et}{2^b}$ **then**
    **return** $\tilde{n}(x,t)$ *(heavy hitter)*
**else**
    Compute $\hat{n}(x,t)$ using (3).
    **return** $\hat{n}(x,t)$ *(interpolate)*
**end if**

---

## 4 Beyond Sketches

Often we want to estimate the likelihood of a sequence of symbols, e.g. *probabilistic graphical models*. Unfortunately it is undesirable to insert long sequences directly into the Count-Min sketch. The reason for this is that, as stated in Theorem 1, the error in approximating the counts is an *absolute* error rather than being relative to the key frequency. While this is not a problem for relatively frequently occurring terms (provided that we know that they are frequent), the error may not be acceptable for low probability (and hence infrequently occurring) keys. We resort to tools from undirected graphical models to obtain $O(1)$ estimates for item frequencies of more complex objects.

**Cliques** For sequences we use a Markovian approximation. That is, we model e.g. trigrams in terms of a product of unigrams or in terms of a chain of bigrams. For concreteness consider modeling the string 'abc':

$$p(abc) \approx p(a) \cdot p(b) \cdot p(c) \qquad \text{Unigrams} \quad (4)$$

$$p(abc) \approx p(a,b)p(c|b) = \frac{p(a,b)p(b,c)}{p(b)} \qquad \text{Bigrams} \quad (5)$$

Clearly, whenever (4) is exact, it is included in (5) as a special case, since the bigram model subsumes the unigram representation as a special case. In general we may use the Markov model (4) and (5) as a frequency estimate for items that are too rare to track.

Since we only have access to the item frequencies rather than the true probabilities of occurrence it is advisable to use hierarchical smoothing. While it would be most desirable to use a Dirichlet-process style smoothing along the lines of [16], the latter is too costly to apply in real time as it requires considerably more than $O(1)$ computation. Instead we may resort to a simplified approach like backoff smoothing:

$$\hat{p}(a) = \frac{n_a + n_0}{n + Ln_0} \text{ and } \hat{p}(ab) = \frac{n_{ab} + n_1 \hat{p}(a)\hat{p}(b)}{n + n_1}. \quad (6)$$

Here $n_0$ and $n_1$ are parameters which describe the amount of backoff smoothing we employ to estimate the joint as a product of the marginals. More generally, we may use Good-Turing or Kneser-Ney [10] smoothing. Eq. (5) can be extended as follows:

**Theorem 6** *Assume that $x$ has an independence structure that can be expressed as an undirected graph $G(V, E)$ in the sense of an undirected graphical model (here $V$ denotes the vertices associated with the coordinates of $x$ and $E$ denotes the edges in the sense of an undirected graphical model). Moreover denote by $T$ a junction tree containing $G(V, E)$ as subgraph with a set of cliques $\mathcal{C}$ and a set of separator sets $\mathcal{S}$, then it suffices if we obtain counts for $x_C$ and $x_S$ with $C \in \mathcal{C}$ and $S \in \mathcal{S}$ to generate frequency estimates via*

$$\hat{p}(x) = n^{|\mathcal{S}|-|\mathcal{C}|} \prod_{C \in \mathcal{C}} n_{x_C} \prod_{S \in \mathcal{S}} n_{x_S}^{-1} \quad (7)$$

**Proof** This follows immediately from the Hammersley-Clifford Theorem [2], the fact that a junction tree contains the maximum cliques of an undirected graph, and the fact that empirical frequencies converge to their probabilities. ∎

In this view (4) and (5) are simple applications of a Markov chain. It also provides us with a recipe for selecting and generating extrapolation formulae for more complex structures.

**Maximum Entropy Estimation** A natural choice for estimating joint probabilities from marginals would be to perform a (smoothed) maximum entropy estimate which is then guaranteed to be consistent. However, this is too costly, as it requires at least $O(m)$ operations, where $m$ is the size of the support of the distribution. Such an estimate will lead to

$$p(x|\theta) = \exp\left(\sum_{i=1}^{d} \theta[i, h_i(x)] - g(\theta)\right) \quad (8)$$

This follows directly, e.g. from [8, 1]. The advantage is that after significant computation to obtain $\theta$ we can provide more accurate frequency estimates at the same cost as the original Count-Min sketch. Note that the likelihood of the data stream is given by

$$\prod_t p(x_t|\theta) = \exp\left(\operatorname{tr} M^\top \theta - ng(\theta)\right). \quad (9)$$

Unfortunately computing $g(\theta)$ is very costly — it requires that we have access to the support of the distribution, which would require an auxiliary data structure in its own right, thus obviating the advantages afforded by sketches. This is why we do not pursue this avenue further in the present paper.

## 5 Experiments

### 5.1 Setup

**Code** We implemented the Count-Min sketch algorithms 2, 3 and 4 in the client-server setting, using C++ and ICE (http://www.zeroc.com) middleware. The experiments were run on quad core 2GHz Linux x86 servers with 16GB RAM and Gigabit network connection, and used sketches with 4 hash functions, $2^{23}$ bins, and $2^{11}$ aggregation intervals (amounting to 7 days in 5 minute intervals). For trigram interpolation, we load Wikipedia into a single 12GB sketch with 3 hash functions, $2^{30}$ bins, and no time aggregation.

**Data** We use two datasets for our research. One is a proprietary dataset containing a subset of search queries of five days in May 2011. The second dataset contains the entire textual dump of the English version of Wikipedia as per January 4, 2012 (we removed the XML markup as a preprocessing step). In all cases full statistics of the data were computed using Hadoop. These counts serve as the gold standard for any sketches and allow us to obtain a proper evaluation of the estimated frequency counts.

Our choice of data (see Figure 6) was guided by both the need to build algorithms applicable in an industrial context and by the need to provide reproducible experimental descriptions. Furthermore, both datasets are quite different in terms of the size of their long tail. The smaller query dataset has a very significant component of infrequent terms whereas Wikipedia, largely prose, displays a much lighter tail with a support of only 4.5 Million unique terms.

### 5.2 Performance

In our experiments all three types of sketches were able to consistently handle over 50k inserts per second. For read requests we measured on average 22k requests per second for time aggregation sketch, and 8,5k requests per second for item aggregation and resolution interpolation. Performance is proportional to the network bandwidth: inserts are one-way, asynchronous, and use very short messages, whereas queries require two-way communication and response messages are larger.

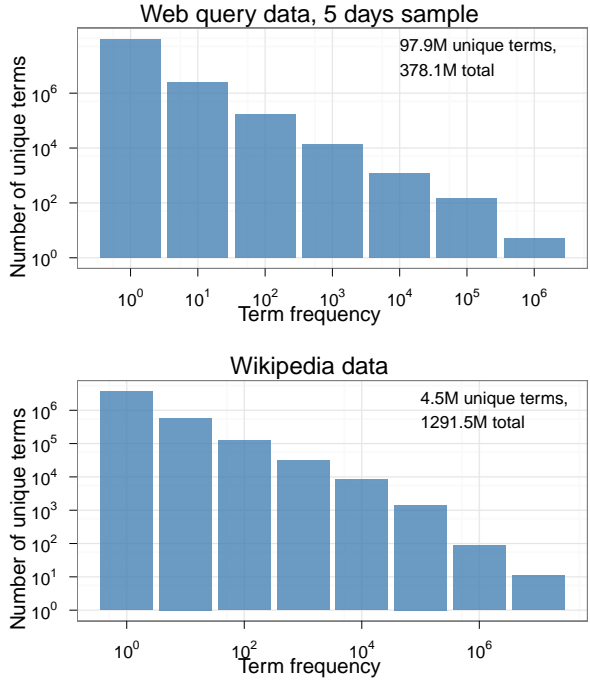

Figure 6: (Top) web query data from 5 days in May 2011; (Bottom) English text in Wikipedia as per January 4, 2012. While both datasets exhibit power-law behavior they differ quite a bit in size and support.

### 5.3 Accuracy

The first set of experiments is to evaluate the accuracy of the sketches under temporal aggregation. Since the Wikipedia dataset constitutes a snapshot in time we only study the accuracy for the query dataset. The Wikipedia dataset is only used to study interpolation. The experimental protocol is as follows:

**Gold Standard** For all experiments we computed the true counts using a batch algorithm on Hadoop. It serves as reference for our estimates.
**Time aggregation** as described in Section 3.1.
**Key aggregation** as described in Section 3.2.
**Interpolation** as described in Section 3.3.
**Absolute deviation** is given by the absolute amount that the sketch deviates from the true counts. In the case of item aggregation sketching estimates are always an overestimate. In all other cases this is not necessarily true. We compute it using $\sum_x |\hat{n}_x - n_x|$.
**Relative deviation** is given by the ratio between the deviation and the estimate, i.e. $\sum_x \frac{|\hat{n}_x - n_x|}{\hat{n}_x}$. Note that this quantity may exceed 1 since we aggregate over all keys.

We are not aware of any sketching algorithm addressing the problem of providing aggregates for arbitrary time intervals. Hence we compare our results to the exact offline aggregation results. Furthermore, we compare to a naive aggregation per time slice and also to a piecewise constant model. One would expect the latter to perform well, given that we are dealing with time-series data which is unlikely to change dramatically in between intervals.

As can be seen in Figure 7, the interpolation algorithm is well competitive relative to algorithms that perform constant interpolation or that reduce item counts. As expected the error increases with the amount of time past. This is the case since we compress data every $2^n$ time steps by a factor of 2. The fact that the error of the 'item aggregation' variant is considerably higher than of the 'time aggregation' algorithm suggests that the item frequency does not vary strongly over time for most items, when compared to the variation between items. Hence the relative frequency of occurrence is generally a better estimate. That said, by using interpolation we combine the best of both worlds and obtain overall good estimates.

To obtain a more detailed estimate we stratify accuracy per item frequency interval, i.e. we stratify by the number of occurrences of the item, as depicted in Figure 8. Not very surprisingly for frequent items, reducing the bit resolution of the sketch is less damaging, as follows directly from Theorem 1, hence for heavy hitters interpolation is on par with item aggregation.

### 5.4 Multigram Interpolation

We show that the factorizing approximation of Section 4 can be used to obtain good estimates of multigrams in both textual documents and query streams. To show that this works for interpolating terms, we approximate trigrams by sketching unigrams and bigrams with a regular Count-Min sketch. We do that in three different ways:

**Unigram approximation** by consuming unigrams and using (4) to estimate trigram probabilities;
**Bigram approximation** by consuming bigrams and unigrams and using (5);
**Trigram sketching** by consuming and querying sketch for trigrams directly.

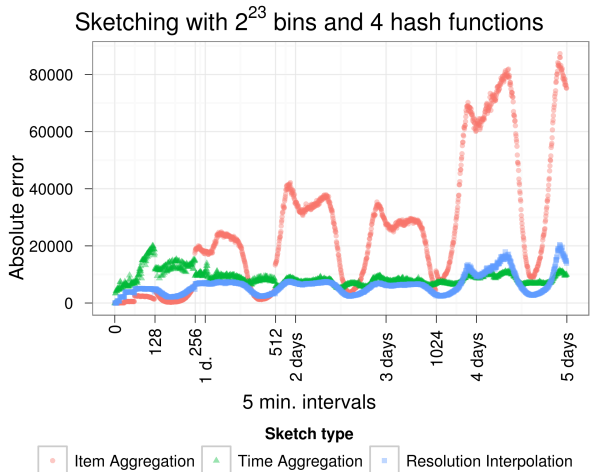

Figure 7: Absolute accuracy $\hat{n} - n$ of three different aggregation algorithms.

|  | Absolute error | Relative error |
|---|---|---|
| Unigram approximation | 24977905.97 | 0.265740 |
| Bigram approximation | 1222160.44 | 0.013002 |
| Trigram sketching | 8352974.60 | 0.088867 |

Table 1: Absolute and relative deviation of trigram approximation models using Wikipedia data.

Table 1 compares the resulting estimates with exact trigram counts. Note that due to the lower number of collisions, our bigram approximation has less than 15% the error of direct aggregation. This is quite significant since it means that in order to store higher order terms direct sketching may not be quite so desirable (possibly with the exception of storing the heaviest hitters directly). It offers a very fast and cheap means of retrieving estimates relative to exact storage.

## 6 Discussion

**Summary** In this paper we presented an algorithm for aggregating statistics on data streams which *retains* temporal information of the items. At its heart we used the Count-Min sketch to obtain a simple yet effective compressed representation of the data. By using time and item aggregation we showed how it is possible to obtain sketches for a more extensive set of query types. Moreover, we showed how the theory of graphical models can be employed to obtain estimates of structured set of covariates.

Our system allows real-time data access in constant time without any need for expensive preprocessing. This makes it an attractive alternative to batch processing systems which, while accurate, are unable to respond to requests without latency and which are therefore less well suited to online data analysis. Experiments demonstrated the excellent performance of the proposed algorithms.

An alternative of our work is to use empirical Laplace transforms to aggregate data. That is, many sketches are amenable to weighted inserts over time. This allows us to obtain sketches of the Laplace-transform counts of a sequence. Decoding then occurs at the client issuing a frequency query to the server. Details of this are the subject of future work.

**Extension to Delayed Updates** In some cases data may arrive with some delay. Here the Count-Min sketch excels due to the fact that it is a *linear* statistic of the data: We can always insert data later into the appropriate (aged) records at a later stage.

A second (and much more significant) advantage is that the additive property of Corollary 2 also applies to sets. Denote by $S(\mathcal{X})$ the data sketch structure given by Algorithms 2, 3, and 4. In this case $S(\mathcal{X} \cup \mathcal{X}') = S(\mathcal{X}) + S(\mathcal{X}')$. Hence we may use MapReduce to carry out sketches on subsets of the data first and then aggregate the terms between mappers. Since everything is linear once the data has been inserted this exhibits perfect scaling properties.

Note that the issue might arise that different subsets of the data span slightly different time ranges. In this case it is important to ensure that all sketches are synchronized to the same time intervals since otherwise aliasing might occur (e.g. a summary of 1 week's data might start on different days on different machines).

**Parallelization** Note that the sketches discussed in the present paper are well amenable to parallelization. In fact, we may use consistent hashing [9] to distribute hash functions and keys over a range of workstations. Details of this strategy are described in [15]. In a nutshell the idea is to send keys to several machines, each of which compute only a single row of the matrix $M$, each of them using a different hash function. The combination of both ideas leads to a powerful enterprise framework for sketching data streams in real time.

**Acknowledgments** This work was supported by the Australian Research Council. We thank George Varghese for inspiring discussions.

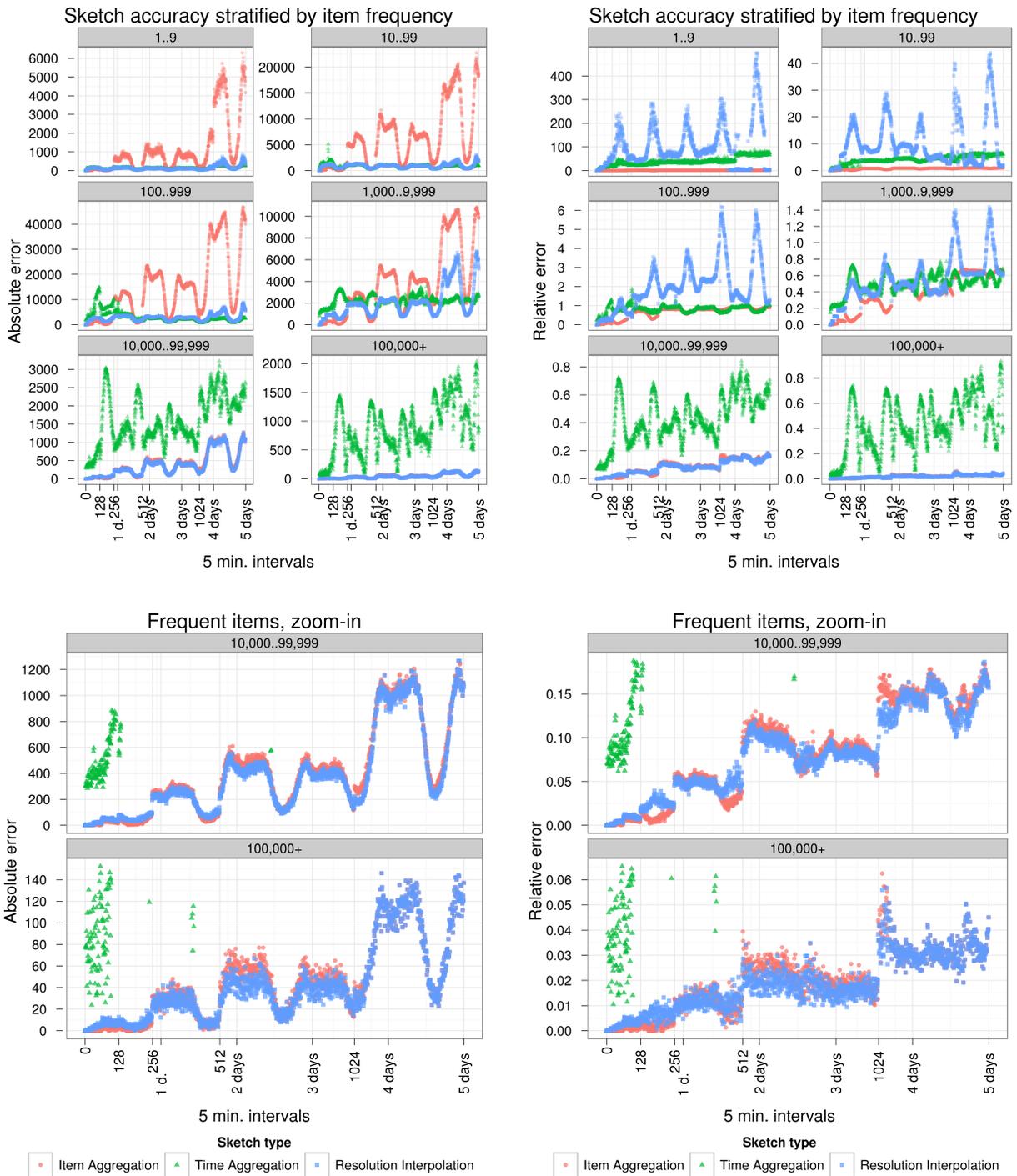

Figure 8: Absolute $\hat{n} - n$ and relative $\frac{\hat{n}-n}{\hat{n}}$ accuracy, stratified by item frequency interval. Left column: absolute error. Right column: relative deviation. Top row: error over time, stratified by frequency of occurrence of items. Bottom row: error for the heaviest hitters.